\documentclass[preprint]{revtex4-2}
\usepackage{xcolor}
\usepackage{hyperref}
\usepackage{amssymb}
\usepackage{amsmath}
\usepackage{graphicx}
\usepackage{enumitem}

\begin{document}
\title{Searching for Scalar Dark Matter via Coupling to Fundamental Constants with Photonic, Atomic and Mechanical Oscillators}
\author{William M. Campbell}
\author{Ben T. McAllister}
\author{Maxim Goryachev}
\affiliation{ARC Centre of Excellence for Engineered Quantum Systems and ARC Centre of Excellence for Dark Matter Particle Physics, Department of Physics, University of Western Australia, 35 Stirling Highway, Crawley, WA 6009, Australia.}
\author{Eugene N. Ivanov}
\author{Michael E. Tobar}
\email{michael.tobar@uwa.edu.au}
\affiliation{ARC Centre of Excellence for Engineered Quantum Systems and ARC Centre of Excellence for Dark Matter Particle Physics, Department of Physics, University of Western Australia, 35 Stirling Highway, Crawley, WA 6009, Australia.}
\date{\today}
\begin{abstract}
We present a way to search for light scalar dark matter (DM), seeking to exploit putative coupling between dark matter scalar fields and fundamental constants, by searching for frequency modulations in direct comparisons between frequency stable oscillators. Specifically we compare a Cryogenic Sapphire Oscillator (CSO), Hydrogen Maser (HM) atomic oscillator and a bulk acoustic wave quartz oscillator (OCXO). This work includes the first calculation of the dependence of acoustic  oscillators on variations of the fundamental constants, and demonstration that they can be a sensitive tool for scalar DM experiments. Results are presented based on 16 days of data in comparisons between the HM and OCXO, and 2 days of comparison between the OCXO and CSO. No evidence of oscillating fundamental constants consistent with a coupling to scalar dark matter is found, and instead limits on the strength of these couplings as a function of the dark matter mass are determined. We constrain the dimensionless coupling constant $d_e$ and combination $|d_{m_e}-d_g|$ across the mass band $4.4\times10^{-19}\lesssim m_\varphi \lesssim 6.8\times10^{-14}\:\text{eV} c^{-2}$, with most sensitive limits $d_e\gtrsim1.59\times10^{-1}$, $|d_{m_e}-dg|\gtrsim6.97\times10^{-1}$. Notably, these limits do not rely on Maximum Reach Analysis (MRA), instead employing the more general coefficient separation technique. This experiment paves the way for future, highly sensitive experiments based on state-of-the-art acoustic oscillators, and we show that these limits can be competitive with the best current MRA-based exclusion limits. 
\end{abstract}
\maketitle
The nature and composition of Dark Matter (DM) is currently one of the most pressing questions in physics. The DM particle composition and interaction properties remain unknown, despite decades of astronomical and cosmological observations indicating that it is the dominant matter component of the Universe. Many different DM candidate particles and corresponding detection experiments have been proposed, for example \cite{Armengaud2012,Aalseth2013,REssig2013,Angloher2016,Gaskins2016,Du2018, Zhong2018,Sikivie2020,gori2020koto,Badurina2020,ORGAN,MADMAX}, but as of yet, no confirmed detections have occurred. In this work, we investigate DM models which add ultra-light scalar fields to the Standard Model (SM). These kinds of models consider non-trivial couplings of putative DM particles, such as the string theory dilaton, and moduli fields to the ordinary SM fields \cite{Green1988,Gasperini2001,Damour2002,Damour2010b,Damour1994}. The introduced scalar DM particle mass is predicted to cause oscillation in some of the fundamental constants of nature, at a frequency equivalent to the DM particle mass, with measurable effects, and potentially inducing violations of Einstein's Equivalence Principle (EEP)~\cite{Damour1994,Damour1990,Stadnik2015,Stadnik2015b}. Consequently, searching for variations in the fundamental constants, or violations of the EEP can be viewed as searches for scalar dark matter candidates. Such experiments are already well developed~\cite{Tobar2013,Tobar2010,Schlam2008,Smith1999,Arvanitaki2015,Tilburg2015,Manley2020,Goryachev2018,Berge2018,savalle2020,Stadnik2016,Stadnik2019} and are motivated from various areas of physics, beyond the search for DM.\\
If such DM-SM couplings exist, oscillations in fundamental constants will frequency modulate various types of clocks and oscillators, measurable via clock  comparison experiments if the oscillators or clocks exhibit different dependences on the fundamental constants. Clock comparisons present a powerful tool for searching for such variations, owing to the high stability of modern frequency standards \cite{Galliou2013, Goryachev2018, Abgrall2016, Goryachev2019}.\\
Confounding these experiments is the fact that the DM particle mass, as well as the strength of its coupling to the SM, is unknown, and only weakly constrained \cite{Primack1988}. As a result there is a large parameter space to search, and experiments are required which search for a range of DM particles masses, corresponding to frequency standards which are stable over different time-scales.\\
In this work we search for variations in the fundamental constants caused by a massive scalar field constituting the local DM halo. We present limits on the possible variation of linear combinations of fundamental constants, thus placing experimental constraints on the coupling of a DM scalar field to the SM. In particular, we contribute new constraints to the relatively unexplored higher mass area of the putative scalar field's parameter space; $4.4\times10^{-19}\lesssim m_\varphi \lesssim 6.8\times10^{-14}\:\text{eV} c^{-2}$. We achieve this by monitoring frequency fluctuations of a quartz crystal bulk acoustic wave oscillator (OCXO), compared against both cryogenic sapphire oscillator (CSO) and hydrogen maser reference clocks (HM), each of which exhibits a different dependence on the fundamental constants.\\
We consider the model of Darmour and Donoghue \cite{Damour2010} as implemented in Ref. \cite{Hees2016}, where $\varphi$ is a dimensionless, massive scalar field with a quadratic self-interaction potential $V(\varphi)=2\frac{c^2}{\hbar^2}m_\varphi^2\varphi^2$ in which the normalisation has been chosen so that $m_\varphi$ has dimensions of mass. This model considers $\varphi$ modifications to terms in the effective action that describes the physics of ground state nuclei. At appropriately low energy scales of $\sim 1$ GeV this effective action will only contain the electron $e$, the up quark $u$ and the down quark $d$ as real particles with interactions mediated by electromagnetic $(A_\mu)$ and gluonic $(A^\text{A}_\mu)$ fields. Weak interactions and heavy quarks are integrated out at this scale, while it is argued in  Ref. \cite{Damour2010} that EEP violation effects linked to the strange quark are relatively small, thus it is ignored here. Each of the five terms described by this effective action can then couple to $\varphi$. For this work, as is common, we consider  a linear couplings for these terms, thus giving a Lagrangian density for scalar field-SM interactions $\mathcal{L}_\text{int}$ as per equation (12) of Ref. \cite{Damour2010}. Linear couplings are often considered to be the most ``simple" and therefore most ``compelling" couplings to the SM, taking the form
\begin{equation}
\begin{split}
\mathcal{L}_{\text{int}}=\varphi\biggl[\frac{d_e}{4\mu_0}(F_{\mu\nu})^2-\frac{d_g\beta_g}{2g_3}(F_{\mu\nu}^A)^2\\
 -c^2\sum_{i=e,u,d}(d_{m_i}+\gamma_{m_i}d_g)m_i\bar{\psi_i}\psi_i\biggr],
\end{split}
\end{equation}
where $F_{\mu\nu}$ is the electromagnetic Faraday tensor, $\mu_0$ the magnetic permeability, $F^A_{\mu\nu}$ is the gluon strength tensor, $g_3$ the QCD gauge coupling, $\beta_g$ is the $\beta$ function for the running of $g_3$, $m_i$ is the mass of the fermions, $\gamma_{m_i}$ is then the anomalous dimension giving the energy running of the masses of the QCD coupled fermions, and $\psi_i$ denotes the fermion spinors. The constants of interest $d_j$ for $j=m_u,m_d,m_e,g,e$ are dimensionless coupling constants that parametrise the scalar field coupling to the SM matter fields, defining the strength of the interaction in a corresponding SM sector, and equivalently, the magnitude of any $\varphi$ dependent oscillations in the corresponding fundamental constants.\\
The introduction of these coupling constants into the interaction Lagrangian density will modify each term such that corresponding fundamental constants will display dependencies on $\varphi$ of the following forms:
\begin{align}
\alpha(\varphi)&=\alpha\left(1+d_e\varphi\right),  \label{eqn:fundvar}\\
m_i(\varphi)&=m_i\left(1+d_{m_i}\varphi\right),\;\;\; \text{ for  }i=e,u,d \label{eqn:fundvar2}\\
\Lambda_\text{QCD}(\varphi)&=\Lambda_\text{QCD}\left(1+d_g\varphi\right) \label{eqn:fundvar3}.
\end{align}
Where $\alpha$ is the fine structure constant, $m_i$ denotes fermion mass (electron ($e$), up ($u$) or down ($d$) quarks), and $\Lambda_\text{QCD}$ represents the QCD mass scale. We also note that the mean quark mass $\hat{m}=(m_u+m_d)/2$  displays a similar dependency on $\varphi$:
\begin{equation}
\hat{m}(\varphi)=\hat{m}\left(1+d_{\hat{m}}\varphi\right),\;\;\text{with}\;\;d_{\hat{m}}=\frac{m_ud_{m_u}+m_dd_{m_d}}{m_u+m_d}.
\end{equation}
The periodic evolution of the scalar field is given by the description in Ref. \cite{Hees2016} which borrows from the cosmological string-theory dilaton model of Ref. \cite{Damour1994};
\begin{equation}\label{eqn:DMevol}
\varphi=\frac{4\pi G\sigma \hbar^2}{m_{\varphi}^2c^6}+\varphi_0\text{cos}(\omega_{\varphi} t+\phi),\;\;\text{with}\;\;\omega_\varphi=\frac{m_{\varphi}c^2}{\hbar}.
\end{equation}
Here; $\sigma = \delta\mathcal{L}_\mathrm{int}/\delta\varphi$ is a source term which is due to the non-minimal coupling between the scalar field and matter. For short time periods ($t\ll 1/H$) this term can be considered a constant. Following the description presented in Ref. \cite{Hees2016}, we identify the scalar field as DM with energy density,
\begin{equation}\label{eqn:density}
\rho_{\text{DM}}=\frac{c^6}{4\pi G \hbar^2}\frac{m_{\varphi}^2\varphi_0^2}{2}.
\end{equation}
We see that for typical values for the local DM density of $\rho_\text{DM}=0.45$ GeV/cm$^3$ \cite{McMillan2011} the amplitude of scalar field oscillations would range from $1.84\times 10^{-13}< \varphi_0 < 1.84\times 10^{-17}$ for $1~\text{mHz} < \omega_{\varphi}/2\pi < 10~\text{Hz}$.
It is thus possible to experimentally probe the coupling of a DM scalar field to matter, by searching for $\varphi$ dependent variations in dimensionless ratios of fundamental constants. This can and has been achieved by making comparison measurements between differing frequency standards whose frequency ratio is dependent on some combination of the dimensionless factors: $m_u/\Lambda_\text{QCD},\: m_d/\Lambda_\text{QCD},\: m_e/\Lambda_\text{QCD}$ and $\alpha$. An experiment of this nature is thus able to probe scalar field couplings linear in  $d_e$ and $d_{m_i}-d_g$, by considering equations~\eqref{eqn:fundvar} to~\eqref{eqn:density}, and the local dark matter parameters.\\
It is often difficult to extract an exclusion limit on individual coupling parameters, owing to these combinations. There are two common methods for extracting information on individual parameters. One is known as `Maximum Reach Analysis' (MRA) in which a series of models are considered where it is assumed that the field in question only couples to one SM sector in each model, thus excluding one non-zero coupling parameter at a time. The other method is to take multiple sets of data with differing dependencies on the constants, so that linear combinations of different parameters can be separated. This is known as `coefficient separation' and is the technique that we employ in this work. By introducing a mechanical resonator, which depends in a different way on the fundamental constants to various photonic and atomic oscillators, we are able to separate coefficients which have not been separated before.\\
\begin{figure}
\centering
\includegraphics[width=8.3cm]{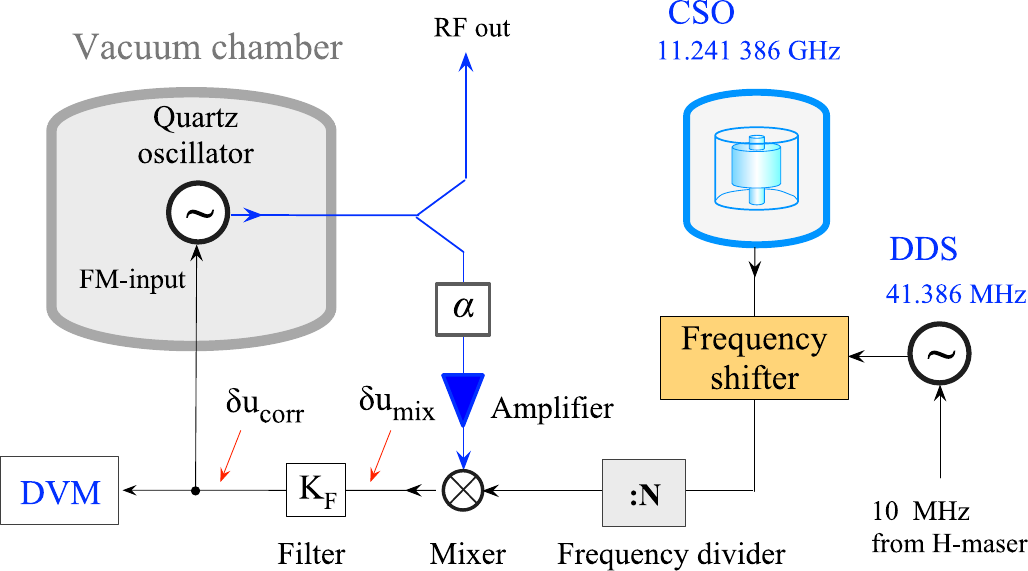}
\caption{Schematic diagram of the initial experimental set up of the OCXO-CSO experiment. An amplifier and attenuator ($\alpha$) are used to prevent injection locking of the quartz oscillator. Injection locking results in spurious bright lines in the spectrum of PLL correction voltage}
\label{fig:exp}
\end{figure}
In one of our experiments we analysed fluctuations of the phase difference $\delta\phi_{21}$ between a 10 MHz NEL Frequency Controls Inc. ultra low phase noise OCXO and a 10 MHz signal synthesised from a microwave CSO \cite{Hartnett2012}. Fig. \ref{fig:exp} shows a schematic diagram of that experiment. Here, the reference 10 MHz signal was synthesized by shifting the CSO frequency by approximately 39 MHz to make it sufficiently (within tens of Hz) close to 11200 MHz before dividing it 112 times. The auxiliary 39 MHz signal was supplied by a Direct Digital Synthesizer (DDS) phase-locked to a 10 MHz CH1-75A active HM. In another similar experiment, the 10 MHz reference signal was supplied straight from the same HM.\\
A Phase-Locked Loop (PLL) was employed to keep mean frequency of the OCXO equal to that of the 10 MHz reference. This was necessary to permit long-term measurements of the phase difference fluctuations $\delta\phi_{21}$.\\
The spectral density of the $\delta\phi_{21}$ was inferred from the spectrum of PLL correction voltage $\delta u_{\textrm{corr}}$ (Fig. \ref{fig:exp}) via the following relationship
\begin{equation}\label{eqn:PLL}
S_{u}(\mathcal{F})=\left|\frac{\gamma}{1+\gamma}\right|^2\left(\frac{\mathcal{F}}{df/du}\right)^2\left(S_{\delta\phi}(\mathcal{F})+S_\phi^{\mathrm{n}/\mathrm{f}}(\mathcal{F})\right),
\end{equation}
where $\mathcal{F}$ denotes the Fourier frequency, $\gamma$ is the loop gain, df/du is the frequency-voltage tuning coefficient of the quartz oscillator, $S_{\delta\phi}$ is the spectral density of phase difference fluctuations and $S_\phi^{\mathrm{n}/\mathrm{f}}$ is the spectral density of the PLL phase noise floor.\\
All parameters of the PLL in Eq. (\ref{eqn:PLL}) can be determined experimentally, thus monitoring the PLL voltage signal $\delta u$ allows us to measure phase noise variations $\varphi_{21}$ that are synchronous with variations in the beat frequency $\delta f_{21}=\delta f_2-\delta f_1$ of two different frequency standards.\\
We now consider how a variation in the fundamental constants will affect the beat frequency between each combination of these frequency standards. To do this we write the dependence of each standard's mode frequencies as a combination of the dimensionless constants $\alpha$, $m_q/\Lambda_\text{QCD}$ and $m_e/m_p\propto m_e/\Lambda_\text{QCD}$ \cite{Tobar2010}. Here we are assuming the constants stated above are functions of $\varphi$ as per eqs. (\ref{eqn:fundvar}) to  (\ref{eqn:fundvar3}), while any other factors contained in the mode frequency dependence are true constants \cite{Turneaure1983}.\\
The derivation of the dependence of the quartz oscillator mode on the fundamental constants is discussed in the appendices and is found to be given by
\begin{equation}\label{eqn:modeQ}
f_{\text{Q}}\propto m_e\alpha^2\sqrt{\frac{m_e}{m_p}}\propto m_e\alpha^2\sqrt{\frac{m_e}{\Lambda_\text{QCD}}}.
\end{equation}
The dependencies of both the CSO and Maser frequencies are given in Appendix A of Ref. \cite{Turneaure1983},as the dependence of the CSO crystal's permittivity on $\alpha$ can be ignored at the frequencies of interest \cite{Tobar2003}. We do not consider here the sensitivity to quark mass through the spin $g$ factor of the HM transition induced by small QCD corrections \cite{Flambaum2004}, as its coefficient is more than an order of magnitude smaller than the other coefficients, and inclusion would cause the other coefficients to lose independence. In future work, to perform the technique of coefficient separation whilst taking into account these corrections, we would require an additional atomic oscillator with a different sensitivity to the quark mass.
\begin{align}
f_{\text{CSO}}&\propto m_e\alpha,\label{eqn:modeCSO}\\
f_{\text{HM}}\propto m_e  \alpha^4 \left(\frac{m_e}{m_p}\right)&\propto m_e  \alpha^4 \left(\frac{m_e}{\Lambda_\text{QCD}}\right).\label{eqn:modeHM}
\end{align}
By normalizing variations in beat frequency with respect to the shared carrier frequency $f_0=10$ MHz, we have, for the quartz against the CSO:
\begin{subequations}\label{eqn:var}
\begin{equation}
\begin{split}
\frac{\delta f_\text{CSO}-\delta f_\text{Q}}{f_0}=-\frac{\delta\alpha}{\alpha}-\frac{1}{2}\left(\frac{\delta m_e}{m_e}-\frac{\delta \Lambda_\text{QCD}}{ \Lambda_\text{QCD}}\right)\\
=-\left[d_e+\frac{1}{2}\left(d_{m_e}-d_{g} \right) \right] \varphi_0,
\end{split}
\end{equation}
\begin{equation}
\begin{split}
\frac{\delta f_\text{HM}-\delta f_\text{Q}}{f_0}=2\frac{\delta\alpha}{\alpha}+\frac{1}{2}\left(\frac{\delta m_e}{m_e}-\frac{\delta \Lambda_\text{QCD}}{ \Lambda_\text{QCD}}\right)\\
=\left[2d_e+\frac{1}{2}\left(d_{m_e}-d_{g} \right) \right] \varphi_0.
\end{split}
\end{equation}
\end{subequations}
Where we have used eqs. (\ref{eqn:fundvar}) to (\ref{eqn:fundvar3}), and $\varphi_0$ has been given by eq. (\ref{eqn:density}).\\
We obtained two initial datasets as per the procedure outlined above. For the CSO-OCXO comparison, data collection took place over just two days, whilst the HM-OCXO data was taken continuously over 16 days. For both datasets, PLL correction voltage time series were collected with a sampling rate of 2.2 Hz. The spectral density of fractional frequency noise; $S_y$, in units of $1/\text{Hz}$, was determined by taking the Fourier Transform (FT) of the beat frequency fluctuations $\delta f_{21}$ and normalizing by the carrier frequency. The range of analysable Fourier frequencies is then determined by the sampling rate of the measurement apparatus and the total integration time. 
In addition to these two initial datasets, further measurements were made at later times over shorter periods in order to provide further complementary results. The CSO-OCXO and HM-OCXO experiments were sampled again, at a higher rate (33 Hz) for 12 hours in order to exclude large noise sources in the initial data, as well as generating fractional frequency noise data at higher frequencies. This data will be subject to the same DM search analysis in the proceeding section, giving less stringent but complementary results. Finally, a further OCXO-OCXO control experiment which displays zero DM sensitivity was run for 12 hours, in order to characterise spurious systematic noise sources in the main data.\\
\begin{figure}
\centering
\hspace{-0.8cm}
\includegraphics[width=7.0cm]{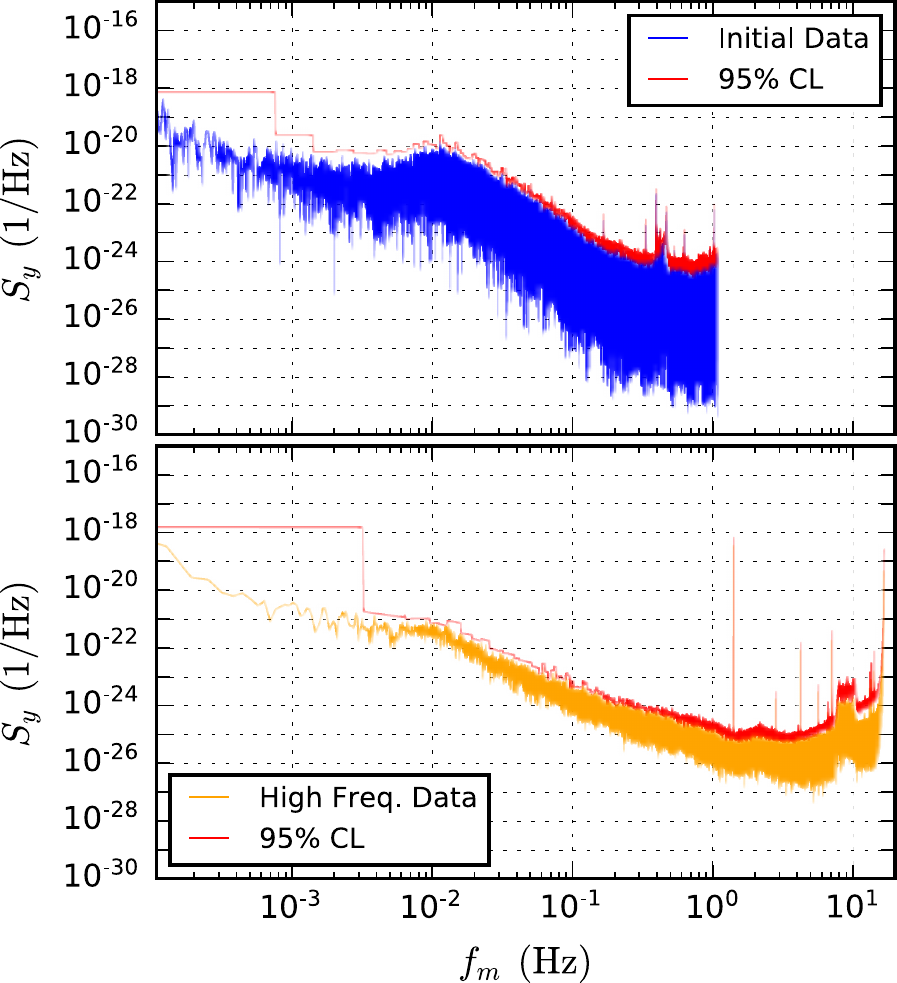}
\caption{Power Spectral Density (PSD) of frequency noise for both initial and later runs of the CSO-OCXO experiment are shown by the blue and orange traces respectively. Also shown in Red is the excluded power to $95\%$ confidence, given by MC simulations. Similar confidence limits were also obtained for the HM-OCXO frequency noise data.}
\label{fig:exec}
\end{figure}
The goal of this analysis is to determine a limit corresponding to the weakest possible scalar field-standard model coupling strength that can be confidently excluded in the case of no detection. The general procedure is as follows. The power spectral density of fractional frequency noise, $S_y$, is searched for large deviations from the mean value at a range of Fourier frequencies, which would correspond to signals consistent with dark matter. A threshold ``cut" value is chosen, and above this cut value any deviations from the mean are considered dark matter candidates. If all such signals can be excluded as either spurious or systematic noise, they are excluded as dark matter candidates, no detection is reported, and we move to derive exclusion limits. A signal size is determined via simulation, which corresponds to the minimum size of a dark matter signal, and thus DM-SM coupling, which we would expect to pass the chosen threshold cut value with 95\% confidence. Given we have excluded all signals above the cut, this simulation-determined signal strength corresponds to our 95\% confidence exclusion limit, as we would have expected a signal of this size to remain in the data, survive the cut, and not be excluded, if it were present.\\
Through the utilised exclusion methods discussed in the appendices, we can exclude all peaks in our data as due to spurious or systematic noise. Furthermore, we are confident that should a signal containing DM characteristics arise; it would fail to be excluded by this analysis, and a strong claim of DM detection could be made.\\
\begin{figure}
\centering
\hspace{-1cm}
\includegraphics[width=7.0cm]{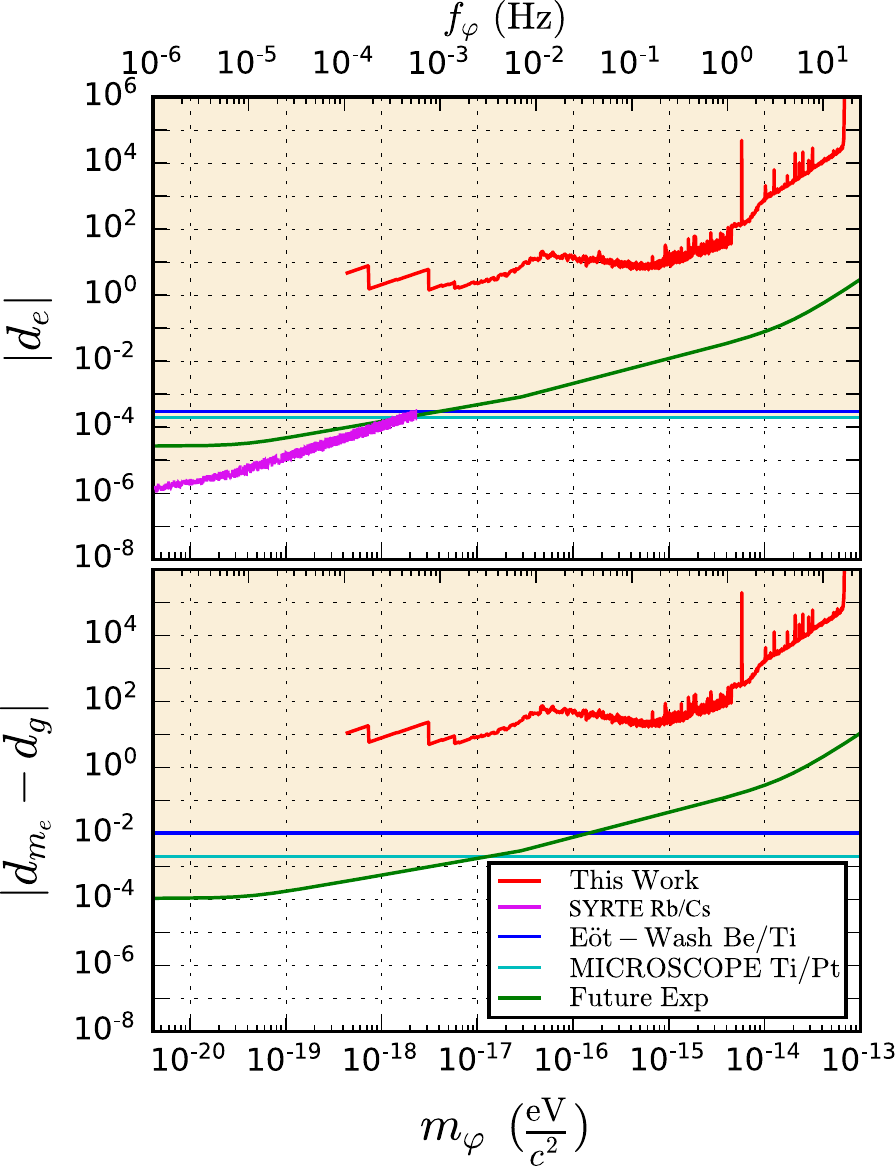}
\caption{Individual limits on coupling constants $|d_e|$ and $|d_{m_e}-d_g|$ derived in this work are presented as the Red trace. This is a combined limit derived from both the initial and higher frequency data. The best known limits in the region, as approximated in Ref. \cite{Hees2018} by performing MRA on results given by the E\"ot-Wash torsion balance EP test \cite{Schlam2008} and MICROSCOPE's WEP test as well as limits on $d_e$ by the SYRTE fountain clock, are shown by the Blue, Cyan and Magenta limits respectfully \cite{Berge2018}. We also further present a projected estimate in Green, for a potential future experiment that would utilise new low phase noise oscillators over a 5 year period.}
\label{fig:newlimits}
\end{figure}
An example of the $95\%$ confidence exclusion limits in terms of frequency fluctuation signal strength for the CSO-OCXO experiment are as shown in Fig. \ref{fig:exec}, similar limits were also computed for the HM-OCXO experiment. The excluded \textit{amplitude} of frequency variation is given by the square root of these limits (when converted back from fractional to absolute frequency), which can then be substituted into Eqs. (\ref{eqn:var}) to give experimental exclusion limits on two different linear combinations of $d_e$, $d_{m_e}$ and $d_g$ (a different set of coupling parameters for each type of oscillator comparison). We then utilise coefficient separation to provide further limits, effectively solving linear equations for $|d_e|$ and $|d_{m_e}-d_g|$. Fig. \ref{fig:newlimits} presents these final exclusion limits derived from both the initial sets of data, and the later higher frequency data, displayed as one combined limit.\\
We note that previous literature limits in this region \cite{Hees2018,Arvanitaki2015} have been approximated by applying MRA to experimental data from EEP and WEP tests. MRA is performed by effectively assuming experimental sensitivity to only one coupling parameter, setting all other parameters to zero. While this method has produced the most competitive limits to date; we note that it is an idealistic estimate, and underpinned by the inherent assumption that any cancellation between multiple parameters is unlikely \cite{Bourgoin2017,Flowers2017}. In deriving our exclusion limits we have considered no case were the coupling to specific SM sectors is temporarily ignored, making them more general. Although such results are obviously inherently less sensitive, they are of slightly different and complimentary significance to those produced by MRA.\\
Recent developments in CSO and OCXO oscillator technologies are giving rise to a new wave of low phase and frequency noise oscillators, far superior to the devices used in this experiment. We have included projected limits in Fig. \ref{fig:newlimits}, as well as frequency stability performance in Fig. \ref{fig:exp}, for a similar hypothetical experiment that searches for variations in CSO-HM and CSO-OCXO frequency differences, using current best-case noise characteristics for such devices from Refs.\cite{Goryachev2018,Ivanov2020,Hmaser}. A 5 year experimental run time in this scenario will achieve general sensitivity limits via coefficient separation that are comparable with those produced by the approximations of MRA, however we note that as sensitivity scales with $T^{1/4}$ the contribution to the improvement of these limits due to longer run times is inferior to that associated with the use of oscillators with better frequency stability. The process for determining these exclusion limits, along with the assumptions made about oscillator performance, are outlined in the appendices. Further improvements to sensitivity could be made by operating the quartz oscillator in a cryogenic environment, where  these oscillators see a boost in quality factor of several orders of magnitude \cite{Goryachev2018,Lo2016,Ivanov2018}. However, due to several practical challenges \cite{Goryachev2012,Goryachev2013}, such a system is yet to be experimentally realized.\\
In conclusion, we present exclusion limits on scalar dark matter coupling to the standard model over several orders of magnitude in dark matter particle mass, based on frequency comparisons of stable oscillators. We exclude parameter space for the coupling constant $d_e$ and combination $d_{m_e}-d_g$. These results represent an improvable, purpose-built experimental contribution to a largely unexplored region of scalar field DM parameter space. The results we have presented are via the coefficient separation method. Although these results are several orders of magnitude less sensitive than those produced by MRA, the fact that these, and any future such limits are not reliant on the assumptions of MRA is a significant strength of this technique. The limits presented here compliment those produced by MRA, as well as other experimental searches in neighbouring regions. Additionally, these results represent the first experimental means to exclude the coupling constant combination $d_{m_e}-d_g$ in isolation, as atomic transition searches display no sensitivity to this combination. Furthermore, we present projected exclusion limits for future iterations of this experimental technique, and show that it has the potential to be competitive with the best MRA limits, without making the same assumptions. We also demonstrate for the first time the power of quartz oscillators as a tool for scalar dark matter detection, and present the derivation of the dependence of the frequencies of such resonators to the fundamental constants.\\
This was funded by the ARC Centre for Excellence for Engineered Quantum Systems, CE170100009, and the ARC Centre for Excellence for Dark Matter particle Physics, CE200100008, as well as ARC grant number DP190100071. We thank Peter Wolf for contributing useful discussions.

\appendix
\section{Bulk Acoustic Wave (BAW) Oscillator Fundamental Constant Dependence}
\label{sec:appendixA}
In determining the dependence of a BAW resonator on the fundamental constants of nature; we begin with the assumption that the BAW phonon resonant frequencies are of the general form
\begin{equation}
f=\frac{nv}{L},
\label{eq:fStatic}
\end{equation}
where $v$ is the speed of sound in the material, $L$ is the relevant length parameter of the resonator, and $n$ is a constant. We will now consider how each of these parameters depends on fundamental constants.\\
The lengths of solids are (to first order, ignoring small relativistic corrections) proportional to the Bohr radius, $a_0$, the characteristic scale for the size of atoms~\cite{AtomicSpacing,GriffithsBohr}. Thus for a solid, such as a BAW resonator,
\begin{equation}
L\propto~a_0\propto~\alpha^{-1}~{m_e}^{-1}.
\label{eq:LChange}
\end{equation}
The speed of sound in a medium is given by
\begin{equation}
v=\sqrt{\frac{K}{\rho}}.
\label{eq:vChange}
\end{equation}
Here K is the relevant elastic modulus for the type of sound wave and material, and  $\rho$ is the density. For the following we consider K to be the Bulk modulus. Given all elastic moduli in solids depend on the balance of the same electrostatic attractive and repulsive forces between atoms, we assume that they depend on the fundamental constants in the same way, and thus sound waves depend in the same way regardless of the modulus chosen. We can rewrite the velocity as
\begin{equation}
v=\sqrt{\frac{K~L^3}{m}},
\label{eq:vChange2}
\end{equation}
by expressing density in terms of $L$ and $m$, the mass. Given that the majority of the mass of a BAW is baryons, the variation of $m$ with constants can be considered proportional to variation of $m_p$, assuming that it is some large multiple of the proton mass.\\
According to~\cite{BM}, for solids with electrostatic inter-atomic bonding, such as quartz
\begin{equation}
K = A {r_0}^{-4},
\label{eq:KChange}
\end{equation}
where $A$ is a (dimensionful) constant relating to the attractive forces between atoms, and ${r_0}$ is the inter-atomic spacing. We assume that, similar to the physical size of the material, the inter-atomic spacing ${r_0}$ depends (to first order, ignoring small relativistic corrections) on the Bohr radius, and so the contributions to the overall dependence from this parameter are known. Dimensionally speaking, we can see that $A$ has dimension $[Nm^2]$, since $K$ has dimensions of $[Nm^{-2}]$ and ${r_0}^{-4}$ has dimension $[m^{-4}]$. According to~\cite{BM} $A$ represents the (electrostatic) attraction term between atoms. Combining these pieces of information, for our purposes, in terms of dependence on fundamental constant, we assume that $A$ is proportional to the Coulomb force, $F_C$ (dimensions [N]), and the square of the characteristic inter-atomic length scale. $F_C$ is given by
\begin{equation}
F_C = \alpha\hbar c \frac{z_1z_2}{r^2},
\end{equation}
where $z_i$ is the atomic number of element $i$, and $r$ is the relevant length scale. For the purpose of the fundamental constant dependence it does not matter which length scale we consider, since we assume that they all vary with fundamental constants in the same way. This ultimately leads us to
\begin{equation}
A\propto F_C \times r^2\propto \alpha.
\end{equation}
This is also the dependence of the Coulomb constant on the fundamental constants, which supports this analysis. This, combined with~\eqref{eq:LChange} and~\eqref{eq:KChange}, yields\\
\begin{equation}
K\propto \alpha^5 {m_e}^{4}.\\
\label{eq:KChange2}
\end{equation}
Substituting~\eqref{eq:vChange2} into~\eqref{eq:fStatic} gives us
\begin{equation}
f=\frac{n\sqrt{\frac{KL^3}{m}}}{L}.
\end{equation}
Finally, substituting~\eqref{eq:KChange2} and~\eqref{eq:LChange}, in terms of fundamental constants (assuming $n$ is a true constant), we arrive at
\begin{equation}
f\propto\sqrt{\frac{KL}{m}}\propto\sqrt{\frac{\alpha^4 {m_e}^3}{m_p}}\propto m_e\alpha^2\sqrt{\frac{m_e}{m_p}}.
\end{equation}
\section{Data Analysis Considerations}
\label{sec:Appendix B}
As discussed in the main body of this work, in order to achieve our goal of determining the limit corresponding to the weakest coupling strength that can be excluded in the case of no detection, we employed a modified version of the power bin search analysis method presented by Daw in Ref. \cite{Daw1998}. Here we present a discussion of the method and the main considerations that were made when applying it to this work.\\
In performing the power bin search method, data acquired over a broadband frequency-binned range is searched for characteristics that correspond to the effect expected to be induced by a DM signal. In the case of \cite{Daw1998} as well as that of our work, the statistic that defines the search is either power excess above the mean level in a single frequency bin, or the total power excess given by the summation of a small number of consecutive bins. Threshold cuts can then be made to the  data to identify candidate bins that contain large power excess potentially due to DM signals. These remaining DM candidates are then systematically eliminated by identifying spurious non-DM related noise signals in the frame of the experiment that co-align at candidate frequencies. If all candidates above the chosen threshold power level can be removed in this way, the threshold can be used to compute a statistical exclusion limit on the DM coupling strength.\\
\begin{figure}\label{fig:candidates}
\centering
\hspace{-1cm}
\includegraphics[width=8.6cm]{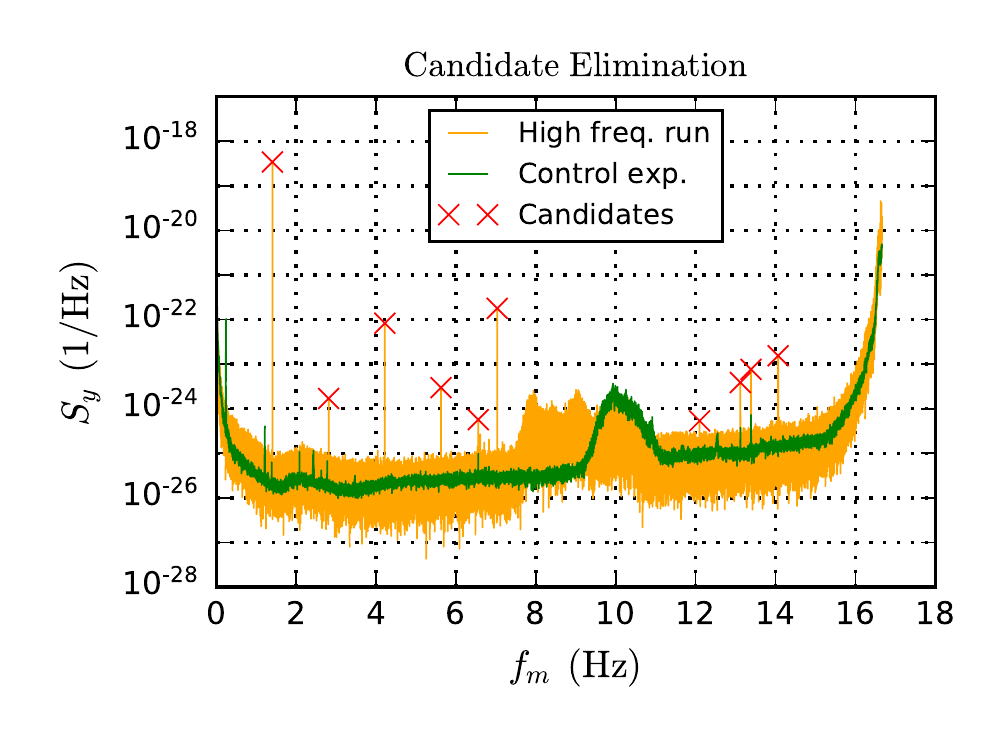}
\caption{Example of the candidate signal characterisation process. Spurious signals in the high frequency data (orange trace) are characterised as noise sources by either comparing with DM insensitive BAW-BAW noise data, or by looking at the signal's width and comparing to that of a characteristic DM signal at the corresponding frequency. In the figure shown, $50\%$ of spurious signals were characterised by the BAW-BAW data, while all signals were also found to be far too broad to be associated with DM.}
\end{figure}
The motivation behind choosing between single or multiple bin search channels is derived from consideration of the DM signal's linewidth (if the DM signal is distributed over a large enough band it will deposit power into multiple neighbouring frequency bins), hence the most optimal search channel is defined by our assumptions of the shape of a DM signal. In this work we are considering a scalar field that constituted the entirety of the local DM halo, thus we assume that this field has been thermalised by its motion in the galactic gravitational potential and therefore any signal, when converted to frequency space, will be defined by a Maxwell Boltzmann distribution of linewidth $10^{-6}\times\omega_\varphi$~\cite{McMillan2011}. This then defines two regimes in the frequency band spanned by our data; the first occurs at low frequencies (DM masses) where the DM linewidth is small and thus its frequency distribution will be completely contained by a single bin. For the data in this regime we conduct a single bin search as described above. The second regime occurs at frequencies where the DM signal's linewidth becomes larger than the width of each frequency bin. For the data in this regime we conduct a 3-bin search, as this provides the most optimal matching of signal linewidth to bin width across this regime of the data set.\\
We also acknowledge that recent theoretical developments \cite{Centers2019} suggest that for integration times less than the coherence time of the DM signal (the first regime in our case), the DM signal exhibits a stochastic fluctuating amplitude, instead of the assumed fixed value $\varphi_0$. For this work, as in the other previous works that we have presented, we assume a deterministic approach where the scalar field amplitude is fixed to an RMS value $\varphi_0$.
\label{sec:appendixc}
\begin{figure}
\hspace{-1cm}
\centering
\includegraphics[width=8.6cm]{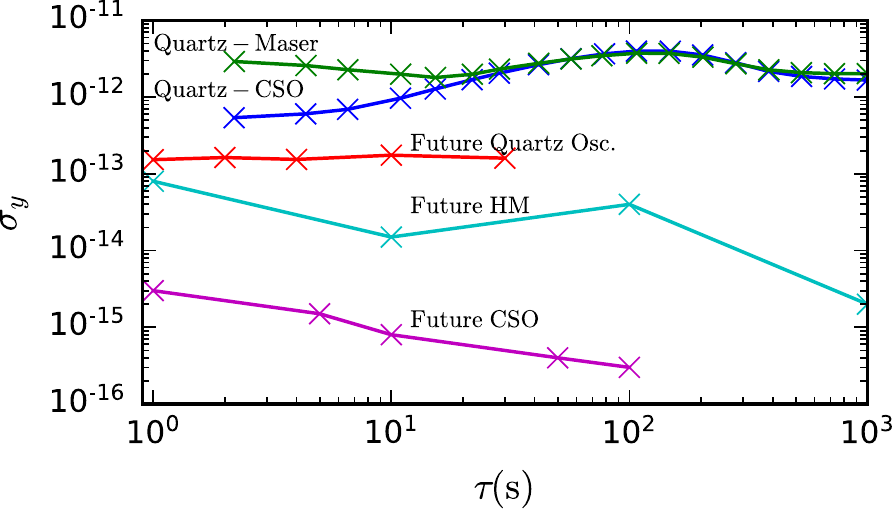}
\caption{Allan variance of the oscillators used in these experiments are shown by the green and blue points. Also included in the red, magenta and cyan points are the expected Allan variances for the oscillators presented in Refs.\cite{Goryachev2018,Ivanov2020,Hmaser}, that we plan to use for future experiments.}
\end{figure}
\begin{figure*}
\centering
\includegraphics[width=16cm]{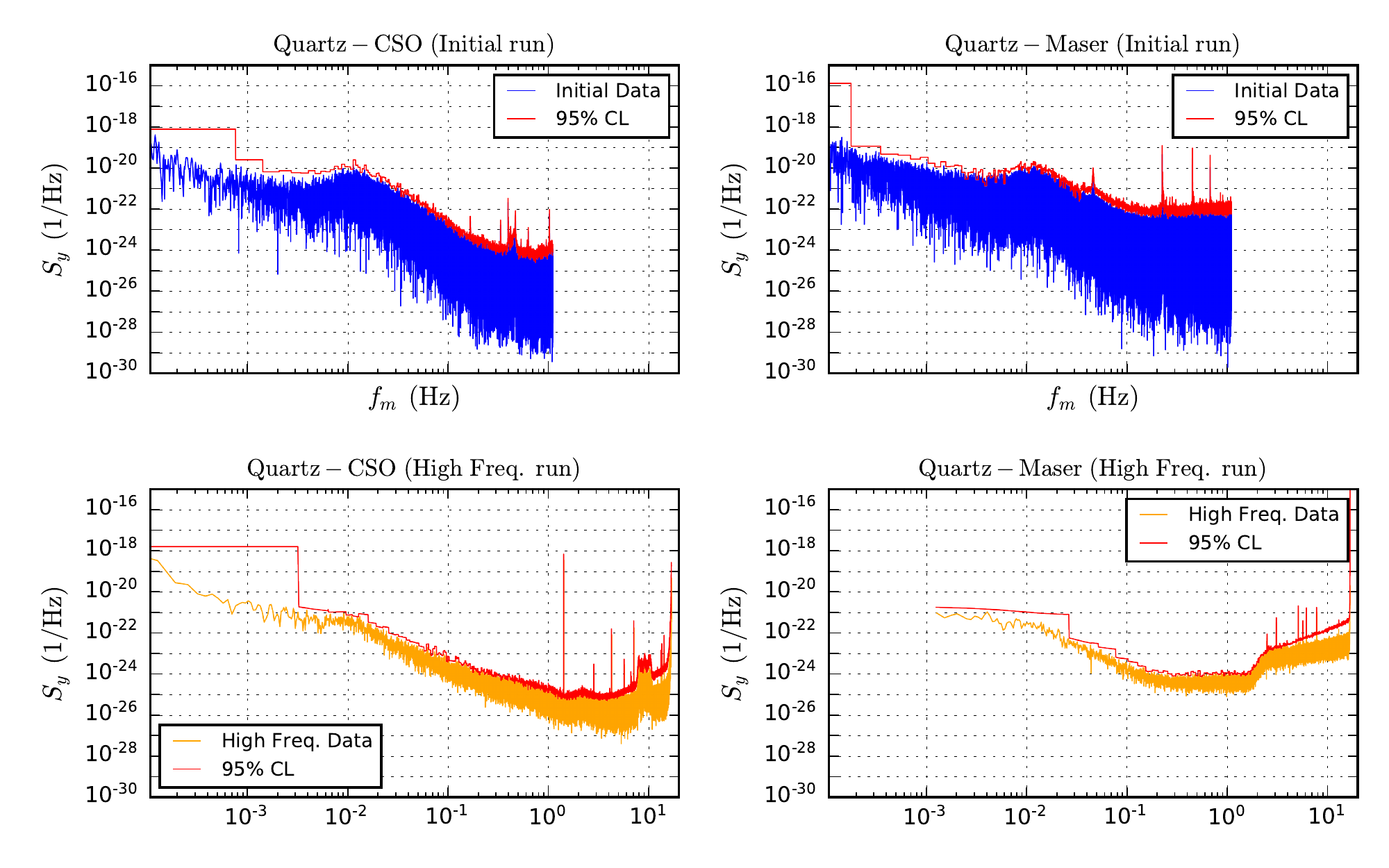}
\caption{PSD of frequency noise for all initial and later runs are shown by the blue and orange traces respectively. Also shown in Red is the excluded power to $95\%$ confidence.}
\end{figure*}
In order to provide exclusion limits across the entirety of our dataset, we must make the claim that no spurious DM-like signals were identified above a threshold cut. In order to support this claim we characterised all large spurious signals seen in the data as non-DM noise by one of three different strategies. The first strategy was to utilise the DM insensitive OCXO-OCXO data to find any large signal peaks at coincident frequencies, thus associating these signals as systematic background noise. Similarly, large signals seen in the 0.1-1 Hz frequency band of the initial data  could be characterised as noise by looking at the DM \textit{sensitive} data from the later higher frequency run, in which these large signals do \textit{not} reoccur. The final method for characterisation is to measure the bandwidth of such signals and compare the measurement to the expected bandwidth of a DM induced effect. As discussed in the preceding section, a DM associated signal would exhibit a bandwidth of $10^{-6}\times f_m$, which for our data corresponds to a width of at most 3 consecutive frequency bins. This allows us to exclude all signals of width $>3$ bins as non-DM associated noise.\\
Using a combination of these three strategies, all spurious signals of large frequency noise excess were able to be characterised as non-DM related noise sources, hence we can claim zero DM detection events and move to determine exclusion limits. In the regions of frequency space where these large signals exist, we simply raise our limits accordingly as to show reduced sensitivity to DM signals in these areas.\\
Should we see any signal that cannot be exclude via the above methods, re-scans would be acquired to confirm its persistence and further higher resolution scans would follow to confirm the signal spectral shape. We are thus confident that should we see a spurious peak of narrow width and large enough magnitude (such that it cannot be excluded by the presented methods) we would be able to make a strong claim of a dark matter signal. We also note that this is a similar procedure for any clock comparison dark matter experiment, such as those referenced in the main text.\\
We utilised Monte Carlo (MC) simulations to convert our threshold frequency noise cut into a $95\%$ confidence limit on the magnitude of frequency shifts induced by a DM scalar field. The simulation's algorithm constructs a multiple bin wide `window' by superimposing a synthesised DM signal injected at some random frequency with a Gaussian-like noise distribution defined by the raw data's statistics. The injected signal's power is then varied until it passes the threshold cut in $95\%$ of simulations. As we interpret our data as containing zero detection events, we define the threshold cut as the power level given by the bin with the highest signal-to-noise ratio (SNR) in each window of the real data, this defines the minimum DM SNR we can exclude. The ensemble of power levels determined by this MC simulation for each window then gives the confidence limit on excluded frequency noise power.\\
In order to estimate sensitivity for a future, improved experiment, we begin by collecting and constructing models of frequency noise in cutting edge oscillators. We take the results presented in Ref. \cite{Goryachev2018} for an estimate of Quartz BAW phase noise, Ref. \cite{Ivanov2020} for an improved CSO estimate, and the values quoted in the technical manual Ref. \cite{Hmaser} for an optimistic HM estimate. Using the gathered frequency noise models we then compute fractional frequency noise power for each oscillator, and add the noise for each set of compared oscillators in quadrature. This gives an estimate of the fractional frequency noise power for a two resonator system, as used in this work.\\
We consider employing a dynamic bin search method (in which we have a varying integration time) on 5 years worth of frequency noise measurements. In order to estimate the smallest fractional frequency noise power which we could exclude from such data, we employ the well known Dicke Radiometer equation 
\begin{equation}
SNR=\frac{S_{y^2_\text{exc}}}{S_{y^2_\text{noise}}}\sqrt{T\times \text{BW}}=\frac{P_{y_\text{exc}}}{S_{y^2_\text{noise}}}\sqrt{\frac{T}{\text{BW}}},
\end{equation}
where $S_{y^2}$ is the spectral density of fractional frequency fluctuations in units of Hz$^{-1}$ with corresponding spectrum of fractional frequency fluctuations $P_y$, $T$ is the total measurement time, and  $\text{BW}$ is the spectral bin width. Thus, setting an $SNR$ goal of $1$ allows us to estimate the lowest fractional frequency noise power we could exclude ($P_{y_\text{exc}}$) in a two resonator experiment. Following the procedure in the main body, taking the square root of this limit gives the excluded amplitude of frequency variation which can then be substituted into an equation similar to that of Eqs. (13) of the main text to give limits on a linear combination coupling constants. We found that the best limits on $|d_e|$ and $|d_{m_e}-d_g|$ would be given by performing coefficient separation on a CSO-HM, CSO-BAW combination, the results of such a projection are presented in the Fig. 4 in the main text.
\bibliography{ScalarDM}
\end{document}